\documentclass[fleqn,usenatbib]{mnras}

\usepackage{newtxtext,newtxmath}

\usepackage[T1]{fontenc}

\DeclareRobustCommand{\VAN}[3]{#2}
\let\VANthebibliography\thebibliography
\def\thebibliography{\DeclareRobustCommand{\VAN}[3]{##3}\VANthebibliography}

\usepackage{graphicx}	
\usepackage{amsmath}	

\usepackage{multirow}
\usepackage{float}
\def\be{\begin{equation}}
\def\ee{\end{equation}}
\def\ba{\begin{eqnarray}}
\def\ea{\end{eqnarray}}

\title[H-Z spectrum in light of tensions]{Return of Harrison-Zeldovich spectrum in light of recent
cosmological tensions}

\author[J.-Q. Jiang et al.]{
Jun-Qian Jiang,$^{1}$\thanks{E-mail: jiangjq2000@gmail.com}
Gen Ye,$^{2}$\thanks{E-mail: ye@lorentz.leidenuniv.nl}
Yun-Song Piao$^{1,3,4,5}$\thanks{E-mail: yspiao@ucas.ac.cn}
\\
$^{1}$School of Physical Sciences, University of Chinese Academy of
Sciences, Beijing 100049, China\\
$^{2}$Institute Lorentz, Leiden
University, PO Box 9506, Leiden 2300 RA, The Netherlands\\
$^{3}$School of Fundamental Physics and Mathematical
    Sciences, Hangzhou Institute for Advanced Study, UCAS, Hangzhou
    310024, China\\
$^{4}$International Center for Theoretical Physics
    Asia-Pacific, Beijing/Hangzhou, China\\
$^{5}$Institute of Theoretical Physics, Chinese
    Academy of Sciences, P.O. Box 2735, Beijing 100190, China
}

\date{Accepted XXX. Received YYY; in original form ZZZ}

\pubyear{2023}

\begin{document}
\label{firstpage}
\pagerange{\pageref{firstpage}--\pageref{lastpage}}
\maketitle

\begin{abstract}
The spectral index $n_s$ of scalar perturbation is the significant initial condition set by inflation theory for our observable Universe. According to Planck results, current constraint is $n_s = 0.965\pm 0.004$, while an exact scale-invariant Harrison-Zeldovich spectrum, i.e. $n_s=1$, has been ruled out at $8.4\sigma$ significance level. However, it is well-known that the standard $\Lambda$CDM model is suffering from the Hubble tension, which is at $\sim 5\sigma$ significance level. This inconsistency likely indicates that the comoving sound horizon at last scattering surface is actually lower than expected, which so seems to be calling for the return of $n_s=1$. Here, in light of recent observations we find strong evidence for a $n_s=1$ Universe. And we show that if so, it would be confirmed conclusively by CMB-S4 experiment.
\end{abstract}




\section{Introduction}
Half a century ago, \citet{Harrison:1969fb}, \citet{Zeldovich:1972zz}, and \citet{Peebles:1970ag}
found that a scale-invariant power law spectrum of primordial
density perturbation is consistent with the crude constraints
available at that time, known as Harrison-Zeldovich spectrum, i.e.
the spectral index $n_s=1$. Recently, the standard $\Lambda$CDM
model has been well inspected with the advent of the era of
precise cosmology. According to Planck 2018 results,
$n_s=1$ has been ruled out at $8.4\sigma$ significant
level\citep{Planck:2018jri}.

However, the case might be not so
simple\citep{DiValentino:2018zjj}. There are some inconsistencies
\citep{Verde:2019ivm,Riess:2019qba,DiValentino:2019qzk,Handley:2019tkm} in standard
$\Lambda$CDM model, which are inspiring a re-interpretation of
currently available data. Recent cosmic microwave background (CMB)
observations showed Hubble constants $H_0\simeq 67$ km/s/Mpc,
However, most local observations give higher values,
e.g. $H_0\simeq 73$ km/s/Mpc using Cepheid-calibrated supernovae \citep{Riess:2021jrx,Riess:2022mme},
although some other measurements, e.g. \citep{Kelly:2023mgv}, show values that are compatible with CMB.
This is
well-known Hubble tension, which has reached $\sim 5 \sigma$
significance level for many observations \citep{Riess:2019qba}, e.g.
\citep{DiValentino:2021izs,Abdalla:2022yfr} for recent
reviews,
see also \citep{Dainotti:2021pqg,Dainotti:2022bzg} for the evolution of $H_0$.
As pointed out in
\citep{Bernal:2016gxb,Aylor:2018drw,Knox:2019rjx}, it is
likely the lower sound horizon $r_s$ at last scattering surface
than expected that results in higher $H_0$, since current
observations require $r_sH_0\sim \text{const.}$.
There have been some attempts to perform $r_s$-independent analysis\citep{Baxter:2020qlr,Philcox:2020xbv,Farren:2021grl,Philcox:2022sgj,ACT:2023kun}, but not robust enough for the solution of Hubble tension\citep{Smith:2022iax}.
It has been found that
if the recombination physics is not modified, in such a
lower-$r_s$ model $n_s$ must be proportionally
raised\citep{Ye:2021nej,Jiang:2022uyg} \be {\delta n_s}\simeq
0.4\left(\frac{\delta H_0}{H_0}\right) \simeq -0.4 \left(\frac{\delta r_s}{r_s}\right),\label{nsr}\ee
which seems to suggest that the complete
resolution $H_0\sim 73$ km/s/Mpc of Hubble tension is pointing
towards a scale-invariant Harrison-Zeldovich spectrum, i.e.$n_s=
1$ ( Here $n_s=1$ refers to a very small region
near $n_s=1$, contrasted with Planck result $n_s = 0.965\pm 0.004$ on
$\Lambda$CDM model), see also \citep{Smith:2022hwi}. Recent
large-scale structure observations is actually not conflicted with
$n_s=1$\citep{Simon:2022adh,Ye:2021iwa}.

As is well known,
the shift of $n_s$ towards $n_s=1$ would have significant and
profound implications for our insight into the inflation theory.
In this work, in light of recent observations, we investigate to
what extent we might live with $n_s=1$.

It is usually thought that the simplest and well-motivated
$V_\text{inf}(\phi)\sim \phi^p$ inflation ($p=2$ for the chaotic
inflation \citep{Linde:1983gd} and $p=2/3,1$ for the monodromy
inflation
\citep{Silverstein:2008sg,McAllister:2008hb,DAmico:2021fhz}) have
been ruled out by Planck+BICEP/Keck \citep{BICEP:2021xfz} due to
their large tensor-to-scalar ratio $r$. However, the case might
not be so. It is possible that initially the inflaton is at
slow-roll region with $N_*\gg 60$, where $N_*$ is the efolds
number before the slow-roll parameter $\epsilon\simeq 1$.
And if
inflation ends prematurely at $N_*- 60$ at which $\epsilon\ll 1$
\citep{Kallosh:2022ggf,Ye:2022efx} by e.g. waterfall instability, we will have
$|n_s-1|=\frac{p/2+1}{N_*}\simeq {\cal O}(0.001)$ (equivalently
$n_s=1$) at CMB observable band, so that both chaotic and
monodromy inflation (also power-law
inflation\citep{DAmico:2022agc}) models will be perfectly
compatible with the recent BICEP/Keck constraint
\citep{Ye:2022efx}, since the uplift to $n_s=1$ markedly lowers
$r\sim |n_s-1|$.


\section{Models and Datasets}

\subsection{Injection of EDE }
The energy injection at $z\simeq z_\text{eq}$ will lead to faster expansion of the universe during recombination, which helps to pull lower the
sound horizon.
A well-known possibility is EDE. In axion-like
EDE\citep{Poulin:2018cxd}, an axion-like potential is
\begin{equation}
V( \phi ) = m^2 f^2 ( 1-\cos({\phi/ f}))^{3},
\end{equation}
while in AdS-EDE\citep{Ye:2020btb}, we have an AdS-like potential
as
\begin{equation}
    V(\phi)=\left\{\begin{array}{ll}
    V_{0}\left(\dfrac{\phi}{M_{\text{Pl}}}\right)^{4}-V_{\text{AdS}} & ,\quad \dfrac{\phi}{M_{\text{Pl}}}<\left(\dfrac{V_{\text{AdS}}}{V_{0}}\right)^{1 / 4} \\
    0 &, \quad\dfrac{\phi}{M_{\text{Pl}}}>\left(\dfrac{V_\text{AdS}}{V_{0}}\right)^{1 / 4}
    \end{array}\right.,
\end{equation}
In corresponding models, $\phi$ starts to roll around redshift
$z_c\sim 3000$. However, their energies must diluted rapidly not
to spoil the fit to CMB. The axion-like EDE achieves it through
oscillation while the AdS-EDE achieves it through the AdS phase
($w>1$).

There are also other possible EDE models, see e.g. \citep{Agrawal:2019lmo,Lin:2019qug,Niedermann:2019olb,Sakstein:2019fmf,Lin:2020jcb,Karwal:2021vpk,Rezazadeh:2022lsf}.

\subsection{Datasets used}

\textbf{Planck}: Planck 2018 low-$\ell$ TT,EE \texttt{Commander}
likeihoods and high-$\ell$ TT,TE,EE \texttt{Plik} likelihoods
\citep{Planck:2018vyg}.\\
\textbf{BAO} (Baryonic Acoustic Oscillations): BOSS DR12\citep{BOSS:2016wmc},
6dF Galaxy Survey \citep{Beutler:2011hx} and Main Galaxy Sample of
SDSS DR7 \citep{Ross:2014qpa}.\\
\textbf{SN} (Supernovae): The Pantheon Type Ia
Supernovae observations\citep{Scolnic:2017caz}.\\
\textbf{R21}: The
measurement of $H_0$ reported by SH0ES \citep{Riess:2021jrx} using
Cepheid-calibrated Type Ia Supernovae is regarded as the Gaussian
constraint on $M_b$.
We use it because it is the typical representative of the local measurements to the Hubble constant, which have been well studied and are consistent with many other local measurements.\\
\textbf{SPT-3G Y1}: The public SPT-3G
likelihood
\footnote{\url{https://github.com/SouthPoleTelescope/spt3g\_y1\_dist}}, which
includes TE and EE spectra within multipoles $300<\ell<3000$
\citep{SPT-3G:2021eoc}.\\
\textbf{ACT DR4}: The marginalized
likelihood \footnote{\url{https://github.com/ACTCollaboration/pyactlike}}
from ACT DR4, which includes TE and EE spectra within multipoles
$326<\ell<4325$ and TT spectra within multipoles $576<\ell<4325$.\\
\textbf{DES Y3}: The measurements of Dark Energy Survey Year 3 on
galaxy clustering and weak lensing \citep{DES:2021wwk}.

\section{Methodology and Results}

As (\ref{nsr}) suggested, a pre-recombination injection, e.g.
well-known early dark energy
(EDE)\citep{Karwal:2016vyq,Poulin:2018cxd}, might be required to
pull lower $r_s$. As examples, we consider the extension of
$n_s=1$ $\Lambda$CDM model, i.e. the $n_s=1$ model with axion-like
EDE \citep{Poulin:2018cxd,Smith:2019ihp} or AdS (anti-de Sitter)
EDE \citep{Ye:2020btb,Jiang:2021bab}.

\begin{figure*}
\centering
\includegraphics[width=.9\linewidth]{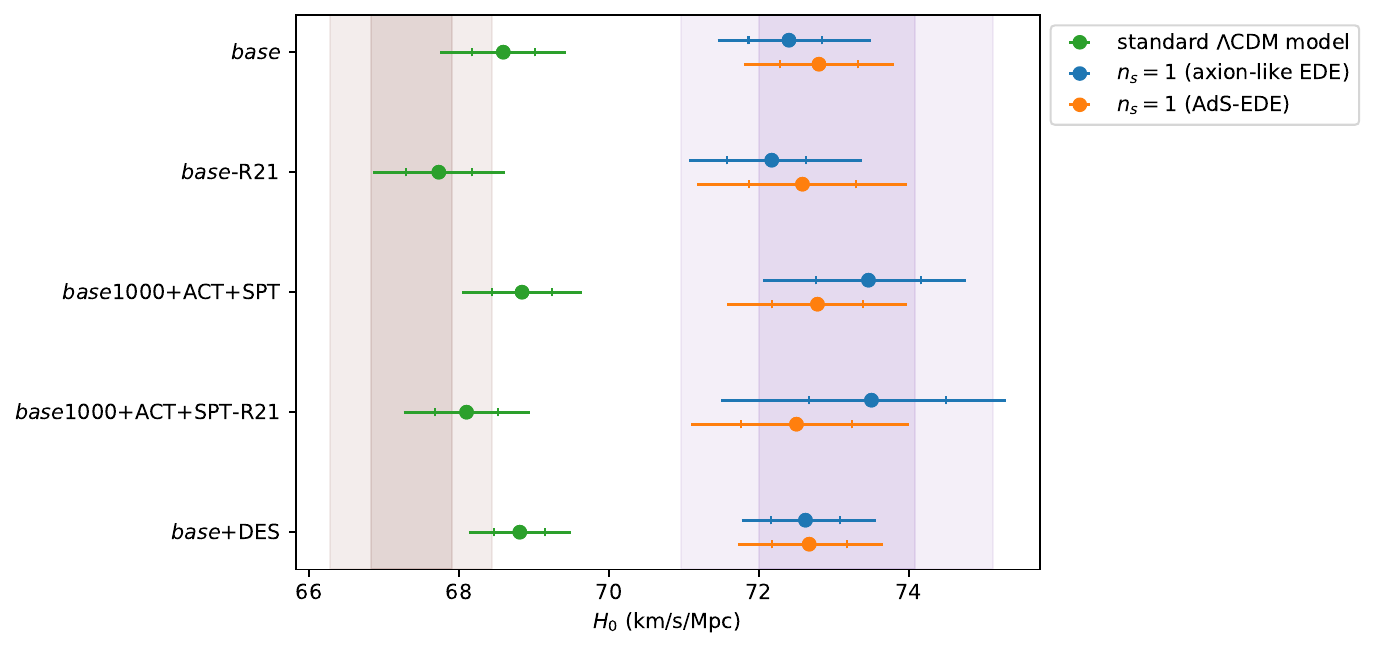}
\caption{The posterior distribution (68\% and 95\% CL region shown as
error bars) of $H_0$ for $n_s=1$ models under different dataset.
The purple band and the brown band represent the local S$H_0$ES
\citep{Riess:2021jrx} constraint (R21) and Planck
2018\citep{Planck:2018vyg} constraint on $H_0$ (68\% CL and 95\%
CL), respectively. } \label{fig:H0}
\end{figure*}

\begin{figure}
\begin{center}
\includegraphics[width=\linewidth]{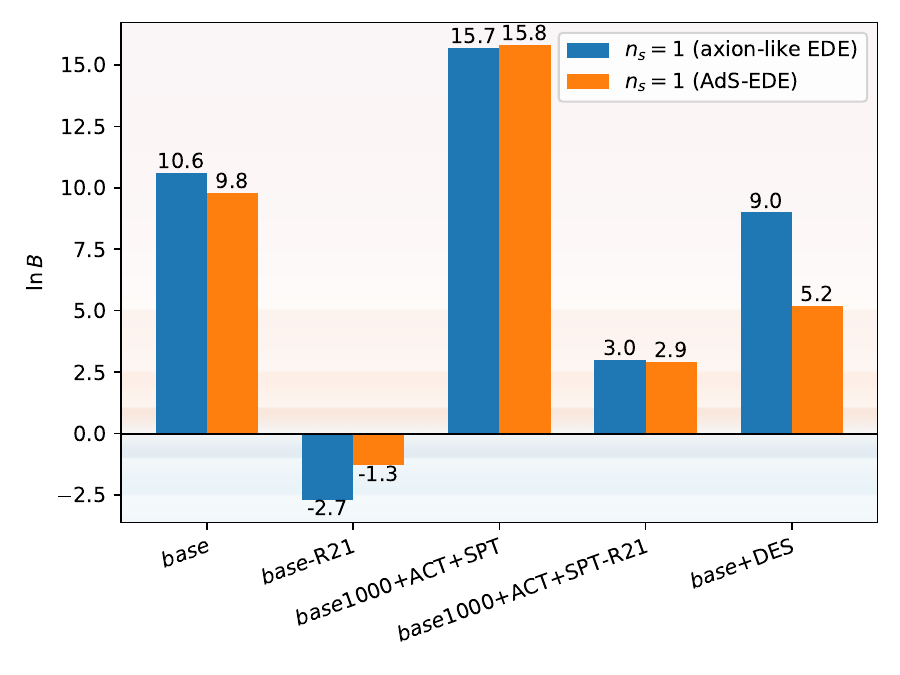}
\end{center}
\caption{Bayes ratio of $n_s=1$ models over $\Lambda$CDM
model for different datasets.
Positive Bayes ratios indicate the preference for $n_s=1$ models.
A revised version of the Jeffreys' scale\citep{Trotta:2008qt} is
also plotted as the background color. } \label{fig:Bayes}
\end{figure}

In what follows we evaluate the Bayesian evidences for such
$n_s=1$ models over the $\Lambda$CDM model. The widely-used
Bayesian ratio is
\begin{equation}
    B = \frac{\mathcal{Z}_{n_s=1\text{ (EDE)}}}{\mathcal{Z}_{\text{standard }\Lambda\text{CDM model}}}.
\label{Bayes}\end{equation}
And the Bayes evidence for model $\mathcal{M}$ is
\begin{equation}
\mathcal{Z}_{\mathcal{M}} \equiv P(\mathbf{D} \mid \mathcal{M})=\int
\mathrm{d} \Theta \mathcal{L}(\mathbf{D} \mid \Theta, \mathcal{M})
\pi(\Theta \mid \mathcal{M}),
\end{equation}
where $\mathcal{L}$ is the likelihood of the data $\mathbf{D}$ for
model $\mathcal{M}$ and parameters $\Theta$, and $\pi$ is the
prior for the model $\mathcal{M}$.
However, Bayesian ratios are prior
dependent. Thus we also consider another prior-independent
quantity, called Suspiciousness \citep{Lemos:2019txn}.
Here, it is calculated based on the Markov chain Monte
Carlo (details are presented in \autoref{sec:MCMC}) results with \citep{Heymans:2020gsg}
\be \ln
S=\left\langle\chi_{n_s=1\text{ (EDE)}}^{2}\right\rangle_\mathcal{P}/2-\left\langle\chi_{\text{standard }\Lambda\text{CDM}}^{2}\right\rangle_\mathcal{P}/2\ee
where $\langle \, \rangle_\mathcal{P}$ means weighted average according to the posterior distribution.
The relevant $p$-value is $p=\int_{(d_{n_s=1} -
d_{\Lambda\text{CDM}})-2 \ln S}^{\infty} \chi_{d}^{2}(x) \mathrm{d} x$,
where $d=2\left(\left\langle(\ln
\mathcal{L})^{2}\right\rangle_\mathcal{P}-\langle\ln
\mathcal{L}\rangle^{2}_\mathcal{P}\right)$ \citep{Handley:2019pqx}.
\textit{As presented in \autoref{fig:H0}, our \textit{base}
dataset is Planck+BAO+SN+R21, and also they are fully compatible
in $n_s=1$ models.} The results of Bayes ratio and Suspiciousness
are shown in \autoref{tab:Suspiciousness} and \autoref{fig:Bayes}.
Both indicate that $n_s=1$ models are favored. \textit{According
to a revised version of the Jeffreys's scale\citep{Trotta:2008qt},
the evidence is very strong, $\ln{B}\simeq 10$ (note that if we
exclude R21, $n_s=1$ models will be not favored, however, such a
compare without Hubble punishment is not fair).}
This conclusion can also be confirmed
by quite negative $\Delta \chi^2$,
see \autoref{tab:chi2}.
As a supplement, we also show the results with
\textit{base1000}+ACT+SPT dataset, which is the
combination of \textit{base} dataset with ACT
DR4\citep{ACT:2020gnv} and SPT-3G Y1\citep{SPT-3G:2021eoc}
observations, while
the small scale part ($\ell>1000$) of Planck TT is cut off, as in \citep{Hill:2021yec,Poulin:2021bjr,LaPosta:2021pgm,Smith:2022hwi,Jiang:2022uyg}.
And it has similar conclusion.

\begin{figure}
\begin{center}
\includegraphics[width=\linewidth]{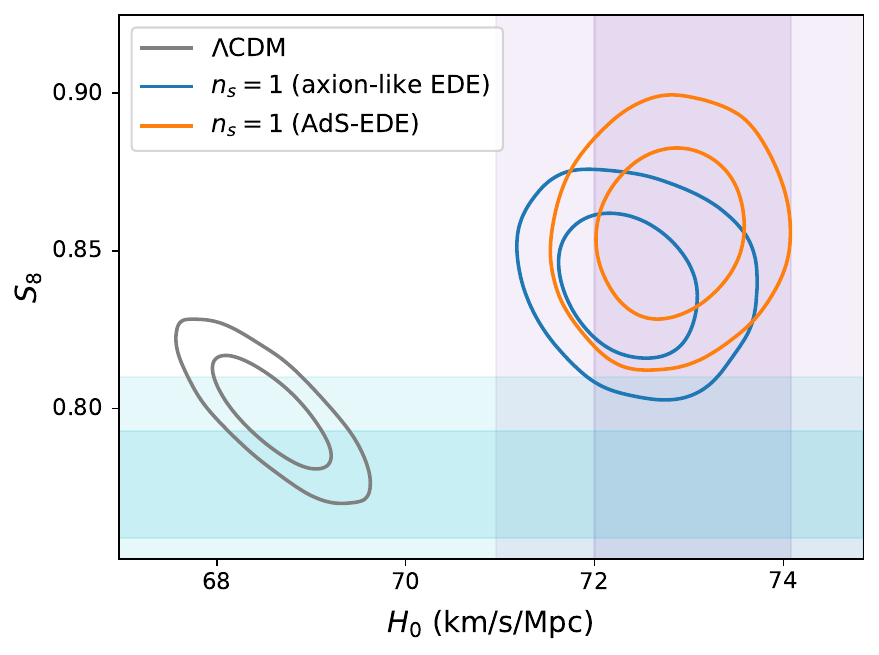}
\end{center}
\caption{ Posterior distribution of model parameters under the \textit{base}
datasets, marginalized to the $S_8$ -
$H_0$ plane. The constraints on $H_0$ from
S$H_0$ES \citep{Riess:2021jrx} (purple band) and $S_8$ from DES-Y3
\citep{DES:2021wwk} (cyan band) are also plotted.  }
\label{fig:tension}
\end{figure}

It is well-known that weak lensing and galaxy clustering
measurements for the growth of structure have $2\sim3\sigma$ level
tension with CMB, quantified as $S_8=\sigma_8
(\Omega_\text{m}/0.3)^{0.5}$, while the injection of EDE will
worsen it \citep{Hill:2020osr,DAmico:2020kxu,Ivanov:2020ril}, see also\citep{Krishnan:2020obg,Nunes:2021ipq}.
As expected, for
$n_s=1$ models we find $S_8\simeq 0.84$, see also
\autoref{fig:tension}. However, with the \textit{base}
dataset+DES-Y3 \citep{DES:2021wwk}, we still find strong evidence
($\ln{B}>5$) for $n_s=1$ models. This might due to that at present
DES-Y3 is not constrained well, although its uncertainty is
already close to other recent observations, such as KiDS-1000
\citep{Heymans:2020gsg}.

\begin{figure}
\begin{center}
\includegraphics[width=\linewidth]{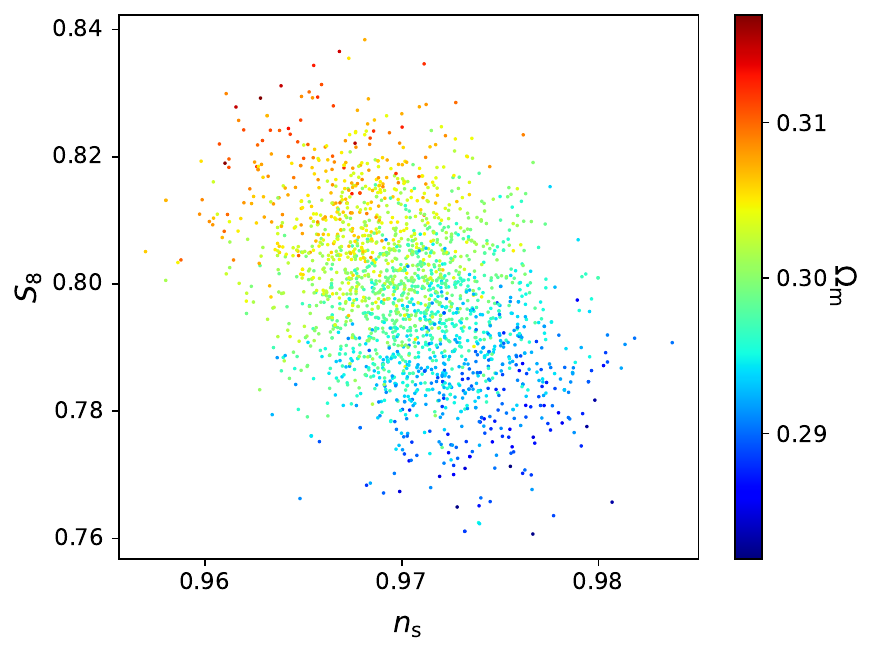}
\end{center}
\caption{ Posterior distribution of the $\Lambda$CDM
parameters under the \textit{base} dataset, marginalized to the
$S_8$ - $n_s$ plane, and the relation with $\Omega_\text{m}$. }
\label{fig:ns-S8-Omegam}
\end{figure}

However, we also need to note that $n_s$ is inversely associated
with $S_8$ through $\Omega_m$, which is clearer in $\Lambda$CDM,
as shown in \autoref{fig:ns-S8-Omegam}. Thus lower $\Omega_m$ can
satisfy $n_s=1$ and a lower $S_8$ simultaneously, but at the cost
of worse fit to BAO+SN. In fact, our $\Omega_m$ is shifted from
$0.2995\pm0.0054$ to $0.2911\pm0.0039$ for
$n_s=1$ model with axion-like EDE.

\begin{table*}
\begin{center}
\caption{Suspiciousness and relevant $p$-values of $n_s=1$
models over $\Lambda$CDM model for different datasets.
Negative Suspiciousness indicate the
preference for $n_s=1$ models.
} \label{tab:Suspiciousness}
\footnotesize
\begin{tabular}{l|cc|cc}
\multirow{2}{*}{Dataset} & \multicolumn{2}{c|}{$n_s=1$ (axion-like EDE)}       & \multicolumn{2}{c}{$n_s=1$ (AdS-EDE)}               \\ \cline{2-5}
                         & \multicolumn{1}{c|}{Suspiciousness} & $p$-value        & \multicolumn{1}{c|}{Suspiciousness} & $p$-value        \\ \hline
\textit{base}          & \multicolumn{1}{c|}{$-13.5$}        & $2\times10^{-7}$ & \multicolumn{1}{c|}{$-8.5$}         & $2\times10^{-5}$ \\
\textit{base}-R21      & \multicolumn{1}{c|}{$0.0022$}       & $0.37$           & \multicolumn{1}{c|}{$6.44$}         & $1$              \\
$base1000$+ACT+SPT     & \multicolumn{1}{c|}{$-17.9$}    & $1\times10^{-9}$ & \multicolumn{1}{c|}{$-13.0$}         & $5\times10^{-7}$              \\
$base1000$+ACT+SPT-R21 & \multicolumn{1}{c|}{$-5.18$}        & $0.00082$        & \multicolumn{1}{c|}{$-0.67$}         & $0.14$           \\
\textit{base}+DES      & \multicolumn{1}{c|}{$-11.54$}       & $7\times10^{-6}$ & \multicolumn{1}{c|}{$-4.33$}        & $6\times10^{-3}$
\end{tabular}
\end{center}
\end{table*}

\begin{table}[t!]
\begin{center}
\caption{$\chi^2_{n_s=1}-\chi^2_{\Lambda\text{CDM}}$ of the
bestfit point for different $n_s=1$ models.} \label{tab:chi2}
\begin{tabular}{l|c|c}
                   & $n_s=1$ (axion-like EDE) & $n_s=1$ (AdS-EDE) \\ \hline
\textit{base}          & $-28.4$                     & $-17.9$              \\
\textit{base}-R21      & $-2.1$                      & $8.6$                \\
$base1000$+ACT+SPT     & $-37.5$                       & $-27.9$               \\
$base1000$+ACT+SPT-R21 & $-11.2$                     & $-3.1$               \\
\textit{base}+DES      & $-21.2$                     & $-7.4$
\end{tabular}
\end{center}
\end{table}

\section{Discussion and Conclusion}
In conclusion, in light of recent observations, we find strong
evidence for a Harrison-Zeldovich Universe. In such $n_s=1$ models
with pre-recombination EDE, we naturally have $H_0\sim 73$
km/s/Mpc without well-known Hubble tension. Though DES-Y3 weakens
the evidence for $n_s=1$ models, it is still critical to check if
the joint analysis of further $S_8$ observations with CMB could
show evidence in support of $n_s=1$, or one might need to rethink
the physics of dark matter.
It should also be noted that the high-redshift halo
abundances observed by James Webb Space Telescope
(JWST) appears to require a higher $n_s$\citep{Klypin:2020tud,Boylan-Kolchin:2022kae}
while the Lyman-alpha forest seems to prefer lower values\citep{Irsic:2017sop,Irsic:2017ixq,Chabanier:2018rga,Palanque-Delabrouille:2019iyz}.
More studies are needed for high redshift and the tension between CMB and large-scale structure observations.

\begin{figure*}
\begin{center}
\includegraphics[width=.9\linewidth]{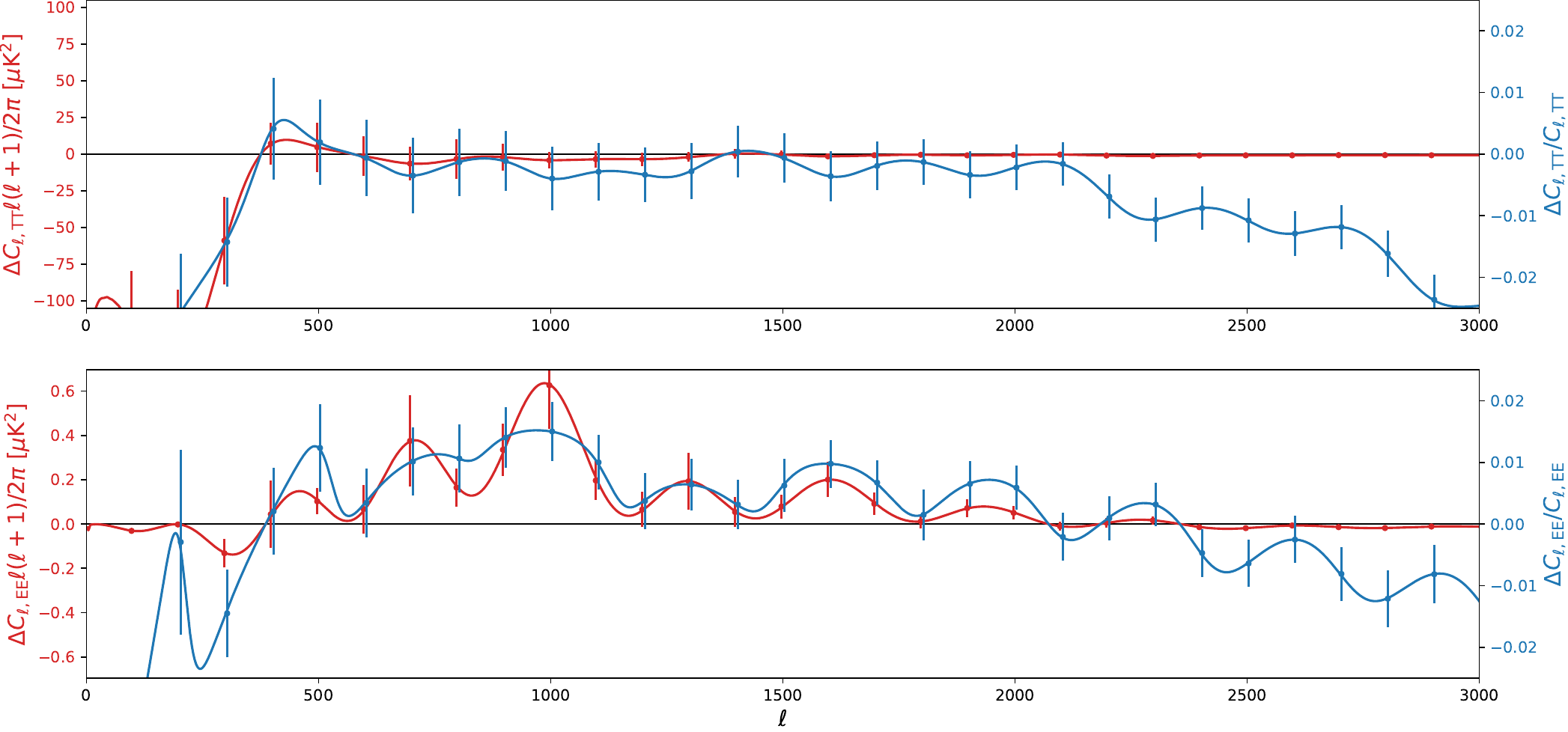}
\end{center}
\caption{ The residuals on the CMB power spectrum of the
bestfit $\Lambda$CDM model, when fitting the mock data assuming a
$n_s=1$ (axion-like EDE) Universe. We also plot the binned CMB-S4
forecasted error bars ($\Delta \ell = 100$). The different colors
represent different scaling shown in the
corresponding vertical axes.} \label{fig:residuals}
\end{figure*}

It is also significant to see, as shown in
\autoref{fig:residuals}, that $n_s=1$ will imprint unique signals
in the CMB spectrum, especially in the small-scale part and
the polarization spectrum, where the impacts of both $n_s=1$ and
EDE are hardly balanced by the shifts of other cosmological
parameters. We make a mock data forecast using CMB-S4
\citep{Abazajian:2019eic}
\footnote{
Regarding the $\Lambda$CDM model as the real universe, we employed
the model parameters from the baseline results (bestfit values) of
Planck 2018 \citep{Planck:2018vyg}:
\begin{equation}
\begin{aligned}
    &H_0 = 67.32 \quad \Omega_b h^2 = 0.022383 \quad \Omega_c h^2 = 0.12011 \\ 
    &\tau = 0.0543 \quad n_s=0.96605 \quad A_s=2.1005\times10^{-9}
\end{aligned}
\end{equation}
Regarding the $n_s=1$ model as the real universe, we employed the
bestfit values from the results of our \textit{base} dataset:
\begin{equation}
\begin{aligned}
    &\log_{10} a_c = -3.858 \quad f_\text{EDE}=0.155 \\
    &H_0 = 72.49 \quad \Omega_b h^2 = 0.023270 \quad \Omega_c h^2 = 0.13439 \\
    &\tau = 0.0626 \quad A_s=2.1550\times10^{-9}
\end{aligned}
\end{equation}
The mock data for CMB-S4 is generated using \url{https://github.com/misharash/cobaya_mock_cmb}\citep{Rashkovetskyi:2021rwg}.
The noise curves for CMB-S4 is taken from \url{http://sns.ias.edu/~jch/S4_190604d_2LAT_Tpol_default_noisecurves.tgz} and details are available at the wiki of CMB-S4: \url{https://cmb-s4.uchicago.edu/wiki/index.php/Survey_Performance_Expectations}.
}
. The results are shown in
\autoref{tab:CMBS4}. The forecast with CMB-S4 indicates that if
the $n_s=1$ model is in reality, i.e.we happened to live in such a
Universe, it would be confirmed conclusively by higher-precision
CMB-S4 experiment.

\begin{table*}
\centering
\caption{ Bayes ratios, Suspiciousness and relevant $p$-values
of $n_s=1$ model with axion-like EDE over the $\Lambda$CDM model
for the CMB-S4 mock data tests.
Positive Bayes ratios and
negative Suspiciousness indicate the preference for $n_s=1$
model.} \label{tab:CMBS4}
\begin{tabular}{l|c|c|c|}
Data                                             & $\ln B$ & Suspiciousness & $p$-value          \\ \hline
Mock data from $\Lambda$CDM model                & $-32.3$ &  $29.3$        & $1$                \\
Mock data from $n_s=1$ (axion-like EDE) model    & $96.7$  & $-86.2$        & $1\times10^{-39}$
\end{tabular}
\end{table*}

\section*{Acknowledgements}

We would like to thank Will Handley, Alessandra Silvestri for valuable comments and suggestions.
This work is supported by the NSFC,
No.12075246 and by the Fundamental Research Funds for the Central
Universities.

\section*{Data Availability}

The data underlying this article will be shared upon request
to the corresponding author(s).



\bibliographystyle{mnras}
\bibliography{example} 




\appendix

\section{Details of MCMC (Markov chain Monte Carlo) analysis}
\label{sec:MCMC}
The cosmological evolution simulations and the MCMC analyses are
performed using \texttt{CLASS} \citep{Blas:2011rf} and
\texttt{cobaya} \citep{Torrado:2020dgo}, respectively, and the
Gelman-Rubin tests for all chains have been converged to
$R-1<0.05$.
The precision settings for \texttt{CLASS} are increased, especially for the calculation of the lensing effect since it has non-negligible effects on ground-based CMB observations, see also the appendix of \citep{Hill:2021yec}.
The injections of EDE are performed with the modified
\texttt{CLASS}: \url{https://github.com/PoulinV/AxiCLASS}
\citep{Smith:2019ihp,Poulin:2018dzj} and
\url{https://github.com/genye00/class_multiscf}. Then we use
\texttt{BOBYQA}\citep{powell2009bobyqa} to find the bestfit points.
The prior range of parameters are summarized in \autoref{tab:priors}
In our calculation, we
fixed AdS depth to the value in \citep{Ye:2020btb}. And we have
confirmed that our results do not strongly depend on the choice of AdS
depth.

\begin{table}
\centering
\caption{The prior range of parameters used in our analysis. Uniform priors are employed for all parameters.} \label{tab:priors}
\begin{tabular}{llc}
\hline \hline
\multicolumn{1}{c}{}            & \multicolumn{1}{c}{Parameter}              & Prior range    \\ \hline
\multirow{6}{*}{$\Lambda$CDM model parameters} & $n_s$ (for standard $\Lambda$CDM model) & $[0.8, 1.2]$ \\
                                & $\log(10^{10} A_\mathrm{s})$               & $[1.61, 3.91]$ \\
                                & $H_0$                                      & $[20, 100]$    \\
                                & $\Omega_\mathrm{b} h^2$                    & $[0.005, 0.1]$ \\
                                & $\Omega_\mathrm{c} h^2$                    & $[0.05, 0.99]$ \\
                                & $\tau_\mathrm{reio}$                       & $[0.01, 0.8]$  \\ \hline
\multirow{3}{*}{EDE parameters} & $\log_{10}(z_c)$                           & $[3, 4]$       \\
                                & $f_\mathrm{EDE}(z_c)$                      & $[0, 0.3]$     \\
                                & $\Theta_\mathrm{ini}$ (for axion-like EDE) & $[0, 3.1]$     \\ \hline
\end{tabular}
\end{table}


\bsp	
\label{lastpage}
\end{document}